\documentstyle[preprint,aps]{revtex}

\begin{document}
\title{Testing Bell's inequality using Aharonov-Casher effect}
\address{Arun Kumar Pati$^*$}
\address{Theoretical Physics Division, 5th Floor, Central Complex}
\address{Bhabha Atomic Research Centre, Mumbai(Bombay)-400 085, INDIA.}
\date{\today}
\maketitle

\begin{abstract}
We  propose  the  Aharonov-Casher  (AC) effect for four entangled
spin-half  particles carrying magnetic moments in the presence of
impenetrable line charge. The four particle  state  undergoes  AC
phase  shift  in two causually disconnected region which can show
up in the correlations between different spin states  of  distant
particles.  This  correlation can violate Bell's inequality, thus
displaying the non-locality for four particle entangled states in
an objective way. \end{abstract}

\vskip 3cm

PACS NO: 03.65.Bz

\vskip 3cm

$email^*$:krsrini@magnum.barc.ernet.in

\newpage

Microphysical  reality  is  certainly  different from the reality
that we are familiar with at a  classical  level.  This  is  best
illustrated  with  Bell's inequality \cite{1} which expresses the
constraint on  the  correlation  experiments  when  locality  and
reality  criterion  are  taken  for  granted.  Quantum mechanical
entangled states violate Bell-like inequalities \cite{2}  showing
the  incompatibility  of  quantum  theory with any classical type
local, realistic hidden variable theory. For entangled states the
principle of separability is not  satisfied,  where  separability
means  that  two  systems  that  belong  to  space-like separated
regions are independent systems. Recent experiments of Aspect  et
al   \cite{3}  using  photon  states  show  that  in  coincidence
experiments the Bell inequality is violated, thus, favouring  the
non-local  aspect  of  the  quantum  theory.  This  non-locality,
however, can not be used to send signals instantaneously  because
that would violate one of the basic principles of relativity.

There is another kind of non-locality in  quantum  theory,  which
comes  from the observable consequences of the existence of gauge
fields. The  wavefunction  of  the  quantum  system  picks  up  a
non-local phase when the particle travels a force free region but
containing   a  non-zero  gauge  field.  The  famous  example  is
Aharonov-Bohm (AB)  effect  \cite{4}  where  a  charged  particle
encircling  a  long solenoid acquires a phase proportional to the
flux enclosed by the solenoid. Similarly, when the role of charge
and magnetic flux is interchanged  we  have  the  Aharonov-Casher
(AC)  effect  \cite{5}.  This effect is some times referred to as
Anadan-Aharonov-Casher (AAC) effect, because Anandan \cite{6} had
independently predicted the same prior to  Aharonov  and  Casher.
Further  using  a  covariant treatment of electrodynamics Anandan
\cite{7} has obtained the same result and the topological  nature
of  the  effect  was  clarified. In AC effect, when a particle of
non-zero magnetic moment goes arround a line charge it acquires a
phase shift proportional to the charge-density. The AB effect has
been clearly verified experimentally \cite{8}  demonstrating  the
non-local  and  topological  aspects  of  the phase shift. The AC
effect in the same spirit can be  regarded  as  a  non-local  and
topological  effect although it is some what different to that of
AB effect \cite{9}. Recently, there has been arguments suggesting
not to call the AC effect as  a  non-local  and  topological  one
\cite{10}  because the effect can be understood using interaction
of magnetic moment with the local Maxwell field.

In  this  paper,  we  would like to demonstrate the AC effect for
four-particle entangled states and using the AC  phase  shift  we
will  show  that  the  probabilities  of  joint  measurements for
distant particles are precisely  the  ones  that  violate  Bell's
inequality.  A  similar  scheme  based on phase transformation of
spin states on distant pairs has been recently proposed by Hacyan
\cite{11}.  The  phase  shifts  are  supposed  to  occur under the
presence of local magnetic fields  in  distant  regions  exerting
force  on  various  particles. However, in the present scheme, we
exploit the topological nature of the  AC  effect  where  distant
particles  experience  no force even though the electric field is
not absent. Thus the violation of the Bell inequality based on AC
effect  would  strengthen  the  non-locality  of quantum world in
contrast to phase shift base on  local  (in  presence  of  force)
interactions.

Following  the  work  of  Hacyan  \cite{11}, consider two pair of
spin-${1 \over 2}$ particles in  their  respective  singlet  states
$|0,0>_{12}$  and  $|0,0>_{23}$.  Let  the  particles $1$ and $2$
carrying magnetic moments $\mu_1$ and $\mu_2$ are emitted from  a
source  in  which  the  total  spin  $S  =  0$ at some point $C$.
Similarly, let  the  particles  $3$  and  $4$  carrying  magnetic
moments  $\mu_3$ and $\mu_4$ are emitted from another source with
total spin $S = 0$ at a point $D$. These particles  are  arranged
in such a way that $1$ and $4$ meet at some point $A$ and $2$ and
$3$  meet  at some point $B$, which are quite far from each other
and in fact  can  be  causually  disconnected.  Imagine  now  the
situation  where  the  spin-1/2  particles  are moving in a plane
(x-y) with their magnetic moments aligned  perpendicular  to  the
plane  of  the  motion  and  there is an impenetrable line charge
aligned perpendicular to the  plane,  also.  The  path  of  these
particles  enclose  the  line charge and each of them experiences
the electric field due to the line charge. But the force on these
particles vanishes identically with proper alignements.

Now, during the propagation of particles $1$  and  $2$  from  the
source  situated at point $C$ the spin states will undergo the AC
phase shift as given by

\begin{eqnarray}
&&|\pm>_1  \rightarrow  exp(\pm i \int_C^A ({\bf E} \times {\mu_1}).d{\bf r} )|\pm>_1,\nonumber\\
&&|\pm>_2  \rightarrow exp( \pm i \int_C^B ({\bf E} \times {\mu_2}).d{\bf r} )|\pm>_2
\end{eqnarray}
and similarly particles  $3$  and  $4$  emitted  from  point  $D$
undergo AC phase shift as is given by

\begin{eqnarray}
&&|\pm>_3 \rightarrow  exp(\pm i \int_D^B ({\bf E} \times \mu_3).d{\bf r} )|\pm>_3, \nonumber\\
&&|\pm>_4 \rightarrow  exp(\pm i \int_D^A ({\bf E} \times \mu_4).d{\bf r} )|\pm>_4
\end{eqnarray}

where $\pm$ sign in phase factors reflect the fact  that  whether
the  magnetic  moment  is  parallel  or  antiparallel to the line
charge. Under the AC phase shifts, the singlet state $|0,0>_{12}$
transforms to
\begin{eqnarray}
&&|0,0>_{12} \rightarrow \cos(\int_C^A ({\bf E} \times \mu_1).d{\bf r} - \int_C^B ({\bf E} \times \mu_2).d{\bf r}) |0,0>_{12} \nonumber\\
&&+ i \sin(\int_C^A ({\bf E} \times \mu_1).d{\bf r} - \int_C^B ({\bf E} \times \mu_2).d{\bf r}) |1,0>_{12},
\end{eqnarray}
and similarly for the state $|0,0>_{34}$ we have
\begin{eqnarray}
&& |0,0>_{34}  \rightarrow  \cos(\int_C^A ({\bf E} \times \mu_3).d{\bf r} - \int_C^B ({\bf E} \times \mu_4).d{\bf r}) |0,0>_{34} \nonumber\\
&& + i \sin(\int_C^A ({\bf E} \times \mu_3).d{\bf r} - \int_C^B ({\bf E} \times \mu_4).d{\bf r}) |1,0>_{34}.
\end{eqnarray}
where $|1,0>$ is the triplet state with total spin  one  magnetic
moment zero.

Thus  introducing  a  line  charge  into the configuration of two
pairs  of  entangled  states  is equivalent to making each pair a
superposition  of  singlet  and  triplet  states  with   distinct
probability  amplitudes.  This  is  a very notable feature of the
phase  transformation  of  the   spin   states   \cite{11}.   The
probability  of  finding  the  singlet  state  $|0,0>_{12}$,  for
example, has changed from  unity  to  $\cos^2(\int_C^A  ({\bf  E}
\times  \mu_1).d{\bf  r}  -  \int_C^B  ({\bf E} \times \mu_2).d{\bf
r})$, which depends on the AC phase shift (depending on the  path
of  the trajectories $C-A$ and $C-B$ ) that the distant particles
undergo.

Before   the  AC  phase shift the state of the combined system is
$|0,0>_{12} \otimes |0,0>_{34}$. The state of the combined system
after the spin states have undergone AC phase shift is  given  by
(see also eq.(3) in \cite{11})

\begin{eqnarray}
&&|\Psi_{TOTAL}> = {1 \over 2} \bigg[-e^{i(\Phi_1 - \Phi_2 -\Phi_3 +\Phi_4)}|1,1>_{14}|1,-1>_{23} \nonumber\\
&&  - e^{-i(\Phi_1 - \Phi_2 -\Phi_3 +\Phi_4)}|1,-1>_{14}|1,1>_{23} \nonumber\\
&& +      \cos(\Phi_A      -      \Phi_B)(|1,0>_{14}|1,0>_{23}     -
|0,0>_{14}|0,0>_{23}) + \nonumber\\
&&i \sin(\Phi_A      -      \Phi_B)(|0,0>_{14}|1,0>_{23}     -
|1,0>_{14}|0,0>_{23}) \bigg]
\end{eqnarray}
where  we  have  denoted  $\Phi_1  =  \int_C^A  ({\bf  E}  \times
\mu_1).d{\bf r}, \Phi_2 = \int_C^B ({\bf E} \times \mu_2).d{\bf r},
\Phi_3 =  \int_D^B  ({\bf  E}  \times  \mu_3).d{\bf  r},  \Phi_4  =
\int_D^A ({\bf E} \times \mu_4).d{\bf r}$ and the relative phases
at  locations  $A$  and  $B$ are $\Phi_A = (\Phi_1 - \Phi_4)$ and
$\Phi_B =(\Phi_2 - \Phi_3)$, respectively.

Then, the argument goes as in \cite{11}.  One  can  calculate  the
joint  probabilities  of  measuring  the  pairs  of  particles at
locations $A$ and $B$ in states of total spin $S = 1$ or $0$ with
$m  =  0$.  For example, the probability of measuring the spin at
$A$ as $1$ and at $B$ as $1$ is given by

\begin{equation}
P(1,1) =  |<\Psi_{TOTAL}|1,0>_{14}|1,0>_{23}|^2  =  {1  \over  4}
\cos^2 (\Phi_A - \Phi_B).
\end{equation}
Similarly we have probabilities $P(0,0) = {1 \over 4} \cos^2 (\Phi_A - \Phi_B),
P(1,0)  =  P(0,1)  =  {1  \over 4} \sin^2 (\Phi_A - \Phi_B)$. The
correlation function defined from the above  joint  probabilities
is given by
\begin{eqnarray}
&& E(\Phi_A,\Phi_B) = {P(1,1) + P(0,0) - P(1,0) - P(0,1) \over P(1,1) + P(0,0) + P(1,0) + P(0,1)} \nonumber\\
&& = \cos2(\Phi_A - \Phi_B).
\end{eqnarray}
This can be shown to violate the inhomogenous Bell's inequality
\begin{equation}
\vert E(\Phi_A,\Phi_B) - E(\Phi_A,\Phi_{B'}) + E(\Phi_{A'},\Phi_B) + E(\Phi_{A'},\Phi_{B'}) \vert \le 2.
\end{equation}
by arranging different points $A,A'$  and   $B,B'$  to  meet  the
particles  $1,4$  and  $2,3$,  respectively. Thus, using AC phase
shift one can check the violation of  Bell's  inequality  thereby
ruling out local realism.

Like  in any other scheme for testing Bell's inequality, here, we
have figured out two locally controlled parameters on  which  the
correlations  depend.  For  example,  in  the  proposed scheme of
Hacyan \cite{11}  the  locally  controlable  parameters  are  the
interaction  time  of  the  particles  with  the  magnetic fields
present at events $A$ and $B$, where one measures the total  spin
of two spin-half particles. In the present scheme the correlation
$E(\Phi_A,\Phi_B)  =  \cos  2(\Phi_A  -  \Phi_B)$  depends on two
parameters, namely the location of the ``meeting points'' $A$ and
$B$,  where  particle  $1$  and  $4$  and  $2$  and   $3$   meet,
respectively. As the relative phase $\Phi_A$ is given by $\Phi_A
= \int_C^A ({\bf E} \times  \mu_1).d{\bf  r}  -  \int_D^A  ({\bf  E}
\times  \mu_4).d{\bf r}$, one can see that the first term depends
on the path length of the trajectory $C-A$ and the second term on
the trajectory $D-A$, respectively. For a given experimental  set
up  the location of the sources $C$ and $D$ are fixed. So one can
only change the location of the ``meeting point'' $A$. Therefore,
$\Phi_A$ can be locally controlled by changing the location  $A$.
Similarly,  we  can  argue that the relative phase shift $\Phi_B$
depends on the path $C-B$ and $D-B$  and  can  be  controlled  by
changing  the location $B$. Therefore, in the proposed scheme the
correlation  $E(\Phi_A,\Phi_B)$  can  be  locally  controlled  by
changing  the  locations  $A$ and $B$. For obtaining different AC
phase shifts the magnetic moments of the particles  $1$  and  $2$
have to be different from the particles $3$ and $4$.

If  the  magnetic  moments  of  all  these  particles  have  same
numerical  value,  then  the  quantum  correlation   reduces   to
$E(\Phi_A,\Phi_B) = \cos 2(\oint ({\bf E} \times \mu).d{\bf r}) =
\cos  (2\mu  \lambda)$,  where  $\mu$  is  the  projection of the
magnetic moment along the line charge and $\lambda$ is the charge
density (charge per unit length) on  the  line.  The  correlation
depends  exactly on the AC phase as in the usual AC effect, where
a single particle in an interference  setup  encircles  the  line
charge  and  acquires  a  phase equal to $\mu \lambda$. Here, the
difference between entangled state AC effect and single  particle
displaying  AC effect is that in former case non of the particles
(either from pair $|0,0>_{12}$ or $|0,0>_{34}$) encircle the line
charge. Yet, the four particle state acquires a AC phase as if  a
single  entity  has  undergone  a motion arround the line charge,
producing the same effect purely because of entanglement.  Since,
there  is  no objective way to associate the total AC phase shift
to one particle or the other of two pairs, we believe that the AC
effect for entangled state is a non-local effect.

To  conclude this note, we have proposed a scheme where two pairs
of entangled states undergo the AC phase shift in the presence of
an impenetrable line charge which  can  be  manifested  in  joint
correlation  measurements  of  spin states that violates standard
Bell's inequality. These phase shifts can be  locally  controlled
by changing the location of the ``meeting points'' of the distant
particles.  In  the  special case, the AC effect for two pairs of
entangled states can be regarded purely as a non-local effect. In
same line of thought one can also explore the non-local geometric
phase \cite{12} concept for entangled states  \cite{13}  to  test
Bell's  inequality  using  two  photon  interference  experiments
\cite{14}.

\end{document}